# Breaks in gamma-ray spectra of distant blazars and transparency of the Universe

G.I. Rubtsov[1] & S.V. Troitsky[1]

[1]*Institute for Nuclear Research of the Russian Academy of Sciences, 60th October Anniversary Prospect 7a, 117312, Moscow, Russia*

**Energetic gamma rays scatter on soft background radiation when propagating through the Universe, producing electron-positron pairs[1]. Gamma rays with energies between 100 GeV and a few TeV interact mostly with infrared background photons whose amount is poorly known experimentally but safely constrained from below by account of the contribution of observed light from known galaxies[2]. The expected opacity of the intergalactic space limits the mean free path of TeV gamma rays to dozens of Megaparsecs. However, TeV photons from numerous more distant sources have been detected[3]. This might be interpreted, in each particular case, in terms of hardening of the emitted spectrum caused by presently unknown mechanisms at work in the sources[4]. Here we show that this interpretation is not supported by the analysis of the ensemble of all observed sources. In the frameworks of an infrared-background model with the lowest opacity[5], we reconstruct the emitted spectra of distant blazars and find that upward spectral breaks appear precisely at those energies where absorption effects are essential. Since these energies are very different for similar sources located at various distances, we conclude that the breaks are artefacts of the incorrect account**

**of absorption and, therefore, the opacity of the Universe for gamma rays is overestimated even in the most conservative model. This implies that some novel physical or astrophysical phenomena should affect long-distance propagation of gamma rays. A scenario in which a part of energetic photons is converted to an inert new particle in the vicinity of the source and reconverts back close to the observer[6,7] does not contradict to our results. This new axion-like particle appears in several extensions of the Standard Model of particle physics[8] and may constitute the dark matter[9].**

Modern instruments of gamma-ray astronomy, namely atmospheric Cerenkov telescopes and the Fermi spaceborn observatory, continue to discover very-high-energy (VHE, energies larger than 100 GeV) gamma radiation from more and more distant sources[4,10,11]. The optical depth of the Universe for VHE photons is large because of the pair production on the extragalactic background light (EBL). One therefore expects additional suppression of the VHE flux from distant sources with respect to that of similar astrophysical objects located relatively nearby. Indeed, the suppression has been observed recently in the Fermi data[12].

However, a precise amount of the suppression is hard to predict theoretically because of the lack of information about the EBL density (for reviews, see e.g. Refs.[13,14]). Indeed, the relevant target photons are mostly infrared, and the extragalactic infrared background can hardly be determined by measurements within the Solar system because of contamination by the so-called Zodiacal light.

Every particular EBL model allows one to reconstruct the emitted gamma-ray spectrum of the source by correcting the absorbed spectrum for the pair-production suppression. Since, for particular distant sources, the observed spectra are reasonably described by a power-law falloff of intensity with energy, the deabsorbed spectra often exhibit hardening, or upward breaks, because the suppression is energy-dependent: the mean free path of GeV photons is of order of the size of the visible part of the Universe while that of TeV photons does not exceed 100 Mpc.

All observed distant sources are blazars, that is belong to a certain class of active galactic nuclei whose relativistic jets point to the observer. While the mechanism of high-energy emission of blazars is disputable, the bulk of their spectral energy distribution is well described by two wide bumps. The low-energy bump is formed by the synchrotron radiation of relativistic electrons while the high-energy one is often attributed to photons which gain their energies by means of the inverse Compton scattering. The overall shape of the distribution is to a large extent determined by the position of the synchrotron peak which sets up the energy scale of the electron population. One often distinguishes two large classes of blazars, namely flat-spectrum radio quasars (FSRQs) with the peak in the radio to infrared and BL Lac type objects (BLLs) with the peak in optical to X-ray bands. However, deabsorbed spectra of distant blazars often exhibit high-energy hardening which is not seen in nearby objects Even a visual inspection of the deabsorbed blazar spectra leads to a conclusion that both the position and the strength of the spectral hardening differ for blazars located at different distances, see Fig. 1 for an example. In what follows, we quantify this observation by a statistical analysis of a large sample of blazars detected in VHE.

To begin with, we formulate three simple hypotheses to be tested which might explain un-usual high-energy breaks in (otherwise smooth and universal) deabsorbed blazar spectra:

(i) the breaks in the emitted spectra have a physical, source-related origin. This most natural explanation suggests that positions and sizes of the breaks should depend on the physical conditions in the sources, which are to a large extent parametrized by the synchrotron peak position, and should not depend on the distance;

(ii) the spectral shape of the assumed EBL density is incorrect, so that the absorption is overestimated for some particular energy. This would result in unphysical breaks always located close to this energy but becoming stronger for more distant sources;

(iii) the overall absorption is uniformly overestimated, for instance, as a result either of incorrect normalization of the assumed EBL density or of ignorance of some other phenomena which decrease the opacity of the Universe. This option would manifest itself in unphysical breaks located at the energy where the absorption becomes essential for a given distance; the breaks should be stronger for more distant sources.

To study the effect quantitatively, we need a sample of blazars located at various distances with fluxes measured beyond the energies where absorption on EBL is significant. We compile the sample from published data of atmospheric Cerenkov telescopes by making use of the TeVCat catalog[3] and supplement it with a set of more distant objects observed by Fermi. The resulting list of objects is given in Table 1 in the Extended Data (ED) section while the selection criteria and the procedure of the sample construction are described in the Supplementary Information (SI) section.

For each of the objects in the sample, we construct a deabsorbed spectrum as described in the SI section. The results we report here are based on the most conservative (i.e., the lowest absorption) EBL model consistent with lower bounds, the "fiducial" model of Gilmore et al.[5]. For comparison, we performed the same procedure for a popular model of Franceschini et al.[15] and obtained similar results.

For the first preliminary test, we fit each spectrum with a power law and, independently, with two power laws with a break, keeping the break position arbitrary. We select these (few) objects for which the fit with a break is better than without it (as determined by the $R^2$ statistics) and the break corresponds to a spectral hardening (it is not so for a few nearby sources). Then we compare the break positions with the values of energy at which the absorption is expected to be significant (namely, with the energy $E_0$ for which the optical depth due to pair production $\tau = 1$). The results are shown in Fig. 2, where we plot the positions of these significant upward breaks versus the redshift of spectral lines $z$, which is the measure of distance at cosmological scales. The breaks happen, on average, at $\log_{10}(E/E_0) = 0.18 \pm 0.32$, statistically consistent with $E = E_0$. This observation favours the option (iii) and suggests that a more statistically solid test of the hypothesis should be performed.

This kind of a test, to which we turn now, constitutes the main part of our study and results in our principal conclusion. We assume now that the position of the break is fixed at $E = E_0$, perform the correspoding fits and study how the strength of the break, determined as the difference $\Delta\Gamma$ between power-law indices below and above the break, depends on $z$. In this ap-proach, we use the information from all sources, even if their individual breaks are not statistically significant. The results are presented in Fig. 3, together with the best-fit approximation, $\Delta\Gamma = (5.08 \pm 0.37) \log_{10} z + (4.05 \pm 0.29)$. For the best fit, $\chi^2 \simeq 9.9$ for 18 degrees of freedom (upper limits in Fig. 3 correspond to the objects for which there is no significant flux measurement above the assumed break; we do not include them in the fit). For the hypothesis of the absence of breaks, $\Delta\Gamma = 0$, the corresponding $\chi_0^2 \simeq 217.5$ (20 d.o.f.); the assumption of a distance-independent break, $\Delta\Gamma = $ const, results in $\chi_C^2 \simeq 201.0$ (19 d.o.f.). These results mean that, speaking in terms of the Gaussian distribution, the absence of distance-dependent spectral hardening is excluded at the 12.4 standard deviations (12.4$\sigma$) level. This is the principal result of our work, and it gives a serious argument in favour of the (iii) hypothesis formulated above.

However, one should worry about potential systematic errors and statistical biases which might affect the result. The sample of blazars we use is by no means complete (which objects are observed by Cerenkov telescopes is determined by the choice of their teams). We discuss a number of potential problems in SI and conclude that it is unlikely that they might affect our result. However, a detailed quantitative study of the biases and systematics is hardly possible without a complete sample of sources, which unfortunately is not expected to be available, at least in relatively near future.

Among previous studies of gamma-ray blazars in the context of the EBL opacity, two groups went beyond the discussion of individual objects and used a sample, like we do here. The first one is the Fermi LAT collaboration[12] which discovered a spectrum suppression by comparison of BLL spectra grouped in large redshift bins. This result was interpreted as the effect of the EBL absorption. Since the bulk of the data they used correspond to the energies for which the opacity is low, this effect was seen in stacked samples only. This result does not contradict to ours because it does not exclude the opacity below the lowest model and even favours it for high energies, cf. Fig. 1 of Ref.[12]. Horns and Meyer[16] concentrated on the sample of blazars detected at very large optical depths, $\tau \geq 2$, and found a 4-sigma evidence for the pair-production anomaly which lead them to conclusions similar to ours. The differences of our approach, which lead to a 12-sigma result, from Ref.[16], are in the choice of the sample (they used only 7 objects for which significant fluxes have been detected for $\tau \geq 2$ and did not use Fermi-LAT data), in the use of simultaneous data only (they stacked data points obtained at different epochs and by different instruments into a single spectrum, which might result in considerable smoothing of the effect because of a high degree of intrinsic variability of blazars, both in the flux and in the spectral index, and of possible relative systematics in the energy scale between the instruments), and the method of the analyzis. They also did not take into account the shift of the mean energy in the bin in the deabsorbed versus observed spectrum, important at large opacities.

Having established that the most conservative EBL model is likely to overestimate the absorption, we turn now to possible implications of the effect we observed. The probability of the pair-production process cannot be questioned: it is calculated in quantum electrodynamics at the center-of-mass energies where no unknown effects are expected to contribute, and has been measured experimentally. The downward change in the amount of target photons is hardly acceptable because the EBL model we use is already the lowest-opacity one and is saturated by lower limits of Ref.[2]. Therefore, one should consider new processes which affect the observed photon flux and are not accounted for in the absorption model which takes into account pair production only. Several models of this kind have been suggested; they invoke either new physical processes or very unusual astrophysical assumptions. A quantitative study of these models in the context of our result is beyond the scope of the present paper (G.R. and S.T., work in progress); however, we give here a brief overview of possible explanations and present very rough estimates for some models. We consider three approaches; as we will see, they can be distinguished by the assumed value of the (presently unknown) intergalactic magnetic field.

Two quite different scenarios invoke similar extensions of the particle-physics Standard model, the so-called axion-like particles (ALPs; see Ref.[8] for a review and a list of references). In external magnetic fields, these hypothetical particles may convert to photons and vice versae. Applied to our problem, this conversion may happen[17] in intergalactic magnetic fields provided they are sufficiently strong ($> 10^{-9}$ G). In this regime, VHE photons convert to ALPs and back during propagation in a way similar to neutrino oscillations. Since the photon state produces pairs but the ALP state does not, this effectively makes the mean free path of a photon longer. In a rough approximation and for the maximal possible photon/ALP mixing, the path becomes longer by a factor of $\sim 3/2$ because there are two photon polarization states and one ALP state in the beam. Using the optical thickness of $2\tau/3$ instead of $\tau$ in our analysis, we obtain the reduction in significance of the distance dependence of breaks from $\sim 12\sigma$ to $\sim 6\sigma$ which suggests that this scenario may not explain the entire observed effect, though a detailed analyzis is required for a firm conclusion.

In the second scenario, intergalactic magnetic fields are assumed to be weaker, $\lesssim 10^{-10}$ G, and therefore insufficient for the photon/ALP transitions which may happen instead in the regions of stronger field around both the source and the observer. The conversion may happen on magnetic fields of galaxies[6], galaxy clusters or superclusters[7]. A qualitative picture of the effect is that $\sim 1/3$ of original photons convert to ALPs near the source and travel unattenuated, to produce $\sim 2/3 \times 1/3 = 2/9$ photons reconverted back in the vicinity of the observer. The rest of the photons attenuate on the EBL in a usual way. Account of this mechanism in our study reduces the effect to $\sim 2\sigma$ thus making it insignificant. We conclude that our result may be explained in this scenario.

The third option[18] does not require new physics beyond the Standard Model; however, it invokes some non-conventional astrophysical assumptions. In this approach, a competitive source of VHE photon *production* along the path from the source to the observer feeds the photon flux which is, in parallel, absorbed in the usual way. These additional photons may be created in interactions of ultra-high-energy cosmic protons, which are assumed, in this model, to be produced in the very same source, with the background radiation. It is not possible to perform a simple rough test of this model in our study because this scenario is necessarily based on rather arbitrary parameters of the hypothetical proton flux. We just note that the assumption that all TeV blazars accelerate the required amount of energetic cosmic protons should be used with a degree of care. Also, the viability of this scenario requires very low values of the intergalactic magnetic fields, $\lesssim 10^{-14}$ G, otherwise charged particles would be deflected and secondary photons would not point back to the source.

While detailed tests of these scenarios versus our results will be presented elsewhere, our preliminary considerations thus favour the ALP convertion/reconvertion scenario[6,7] for the explanation of the effect we observe.

**Acknowledgements** The authors are indebted to Oleg Kalashev for discussions. This work was supported in part by the RFBR grant 13-02-01293 (G.R. and S.T.), by the Dynasty foundation (G.R.), by the grants of the President of Russia MK-1170.2013.2 (G.R.) and NS-2835.2014.2 (G.R. and S.T.). We acknowledge the use of data and software provided by the Fermi Science Support Center.

**Figure 1**  Example of two spectra with breaks: 3C 279 (redshift $z = 0.536$, large points) and PKS J0730-1141 (redshift $z = 1.591$, small points). Both objects are FSRQs and have similar spectral energy distributions in the optically thin part. However, spectral breaks in the gamma-ray band are located at the energies where the absorption becomes significant, different for the two sources. Grey points represent the observed spectrum, dark points are corrected for the EBL absorption with the most conservative model.

**Figure 2**  Positions of individual significant upward breaks versus redshift $z$. The line represents the energy $E_0(z)$ at which the optical depth with respect to the pair production $\tau = 1$.

**Figure 3**  Value $\Delta\Gamma$ of the break in the spectrum deabsorbed with the most conservative model, assumed to happen at $E = E_0$, versus redshift $z$. The line gives the best fit; its slope is non-zero at the $12\sigma$ significance.

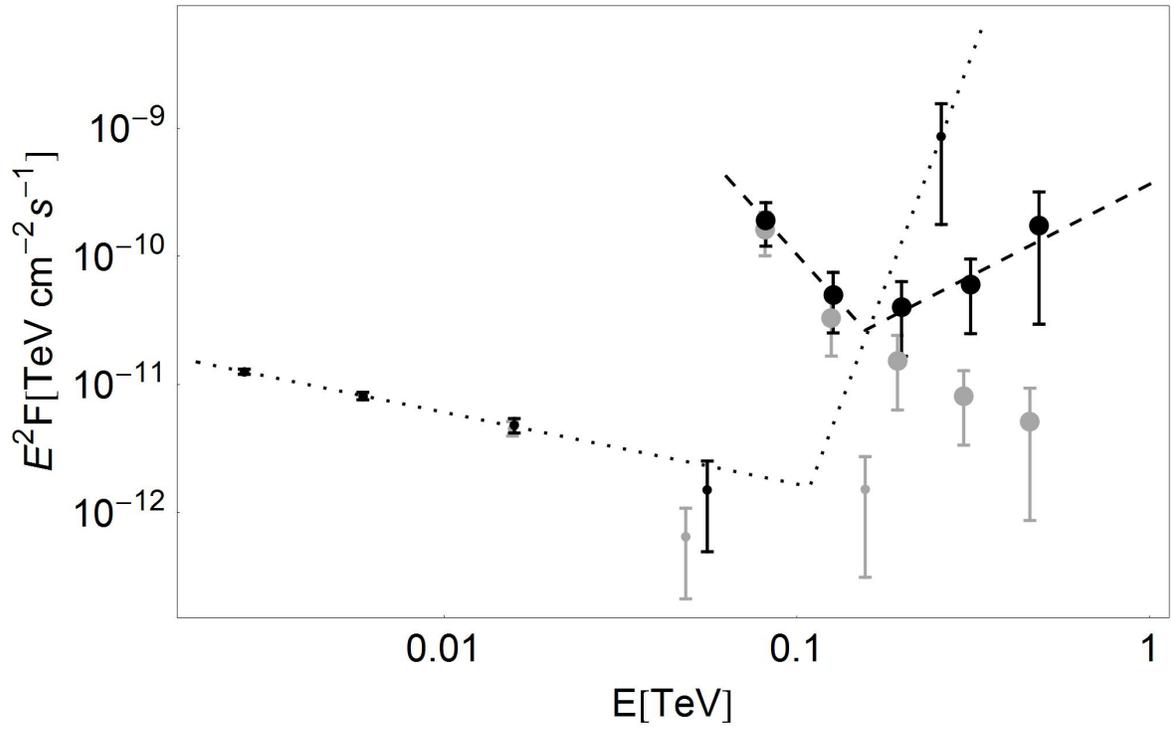

Figure 1:

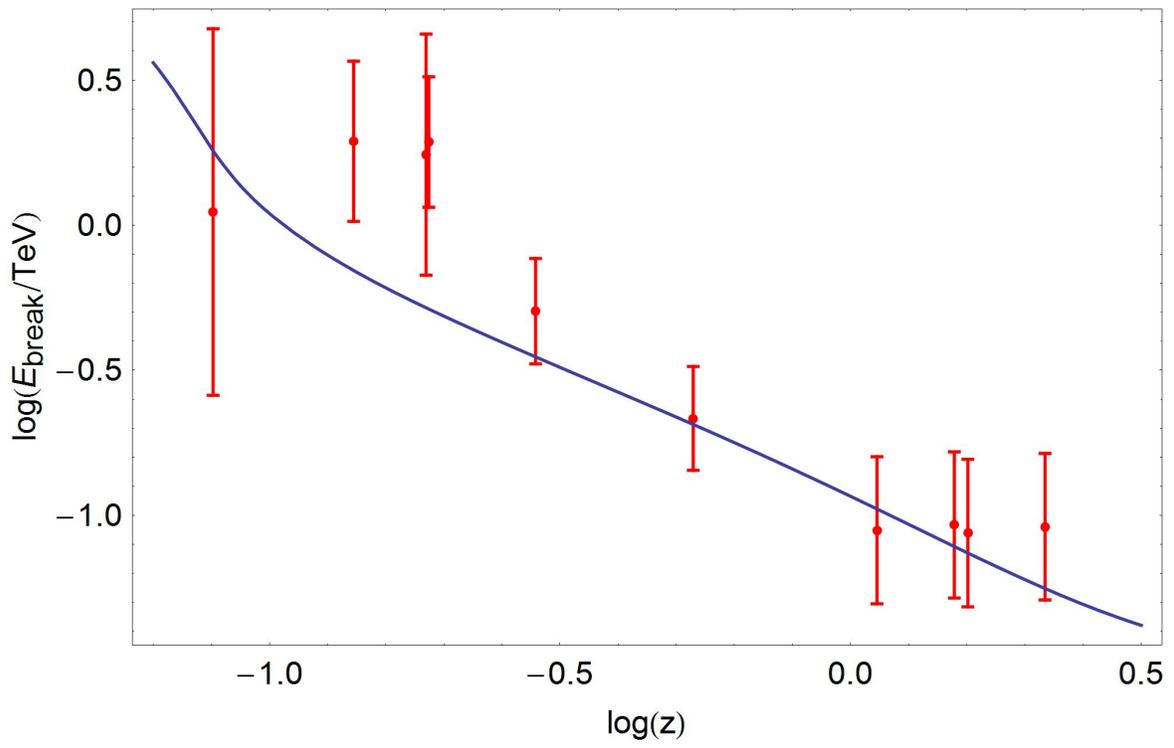

Figure 2:

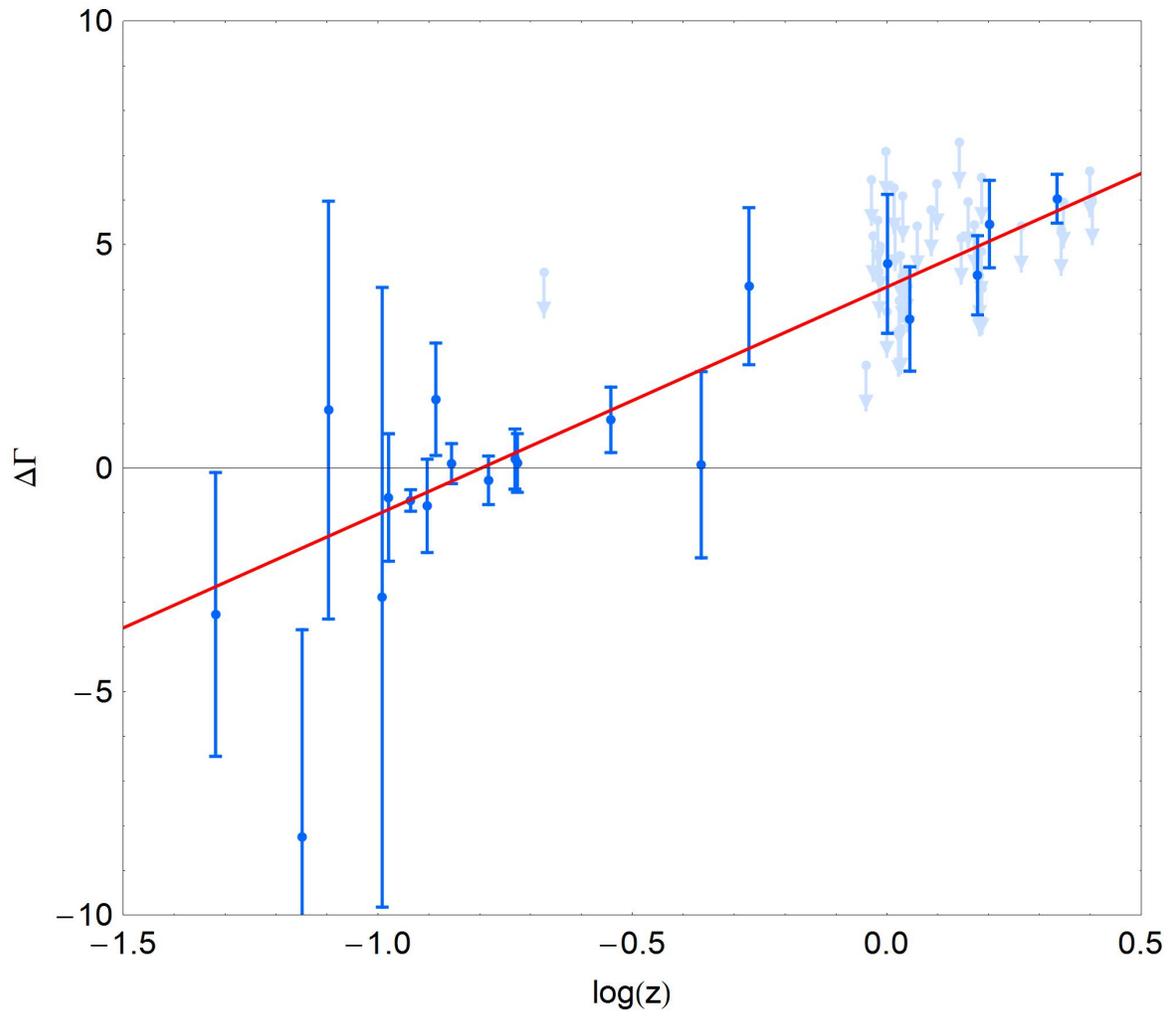

Figure 3:

# Extended Data

| Name | Redshift | Class | Instrument | Reference |
|---:|:---:|:---:|:---:|:---:|
| 1ES 1959+650 | 0.048 | BLL | VERITAS | [19] |
| PKS 2005−489 | 0.071 | BLL | HESS | [20] |
| RGB J0152+017 | 0.080 | BLL | HESS | [21] |
| W Com | 0.102 | BLL | VERITAS | [22] |
| 1ES 1312−423 | 0.105 | BLL | HESS | [23] |
| PKS 2155−304 | 0.116 | BLL | HESS | [24] |
| RGB J0710+591 | 0.125 | BLL | VERITAS | [25] |
| 1ES 1215+303 | 0.130 | BLL | MAGIC | [26] |
| 1ES 0229+200 | 0.140 | BLL | VERITAS | [27] |
| H 2356−309 | 0.165 | BLL | HESS | [28] |
| 1ES 1101−232 | 0.186 | BLL | HESS | [29] |
| 1ES 0347−121 | 0.188 | BLL | HESS | [30] |
| 1ES 0414+009 | 0.287 | BLL | HESS | [31] |
| 4C +21.35 | 0.432 | FSRQ | MAGIC | [32] |
| 3C 279 | 0.536 | FSRQ | MAGIC | [10] |
| PKS 0426−380 | 1.003 | FSRQ | LAT | – |
| PKS 0426−380 | 1.110 | BLL | LAT | – |
| RGB J1448+361 | 1.508 | BLL | LAT | – |
| PKS J0730−1141 | 1.591 | FSRQ | LAT | – |
| B3 1307+433 | 2.156 | BLL | LAT | – |

Table 1: **The list of sources.** The table lists the objects contributed to the main result of the paper, Fig. 3. The sample construction is described in SI. Additional 39 objects which give upper limits only are not listed.

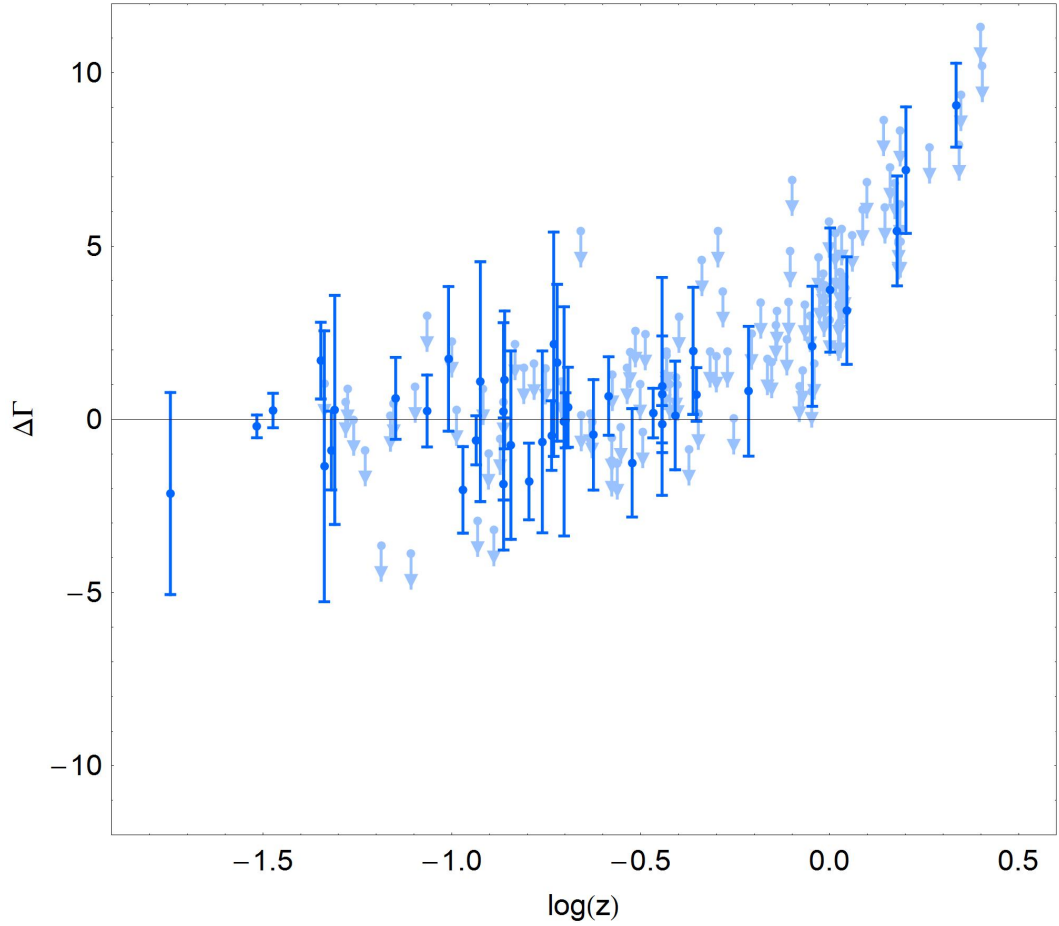

Figure 4: Same as in Fig. 3 but for the break assumed to happen at $E = 100$ GeV, for the extended sample of Fermi-LAT blazars described in the text. The breaks appear for distant objects only, for which $E_0 \sim 100$ GeV.

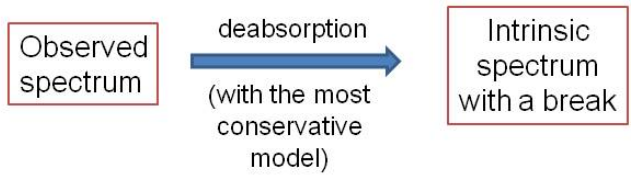
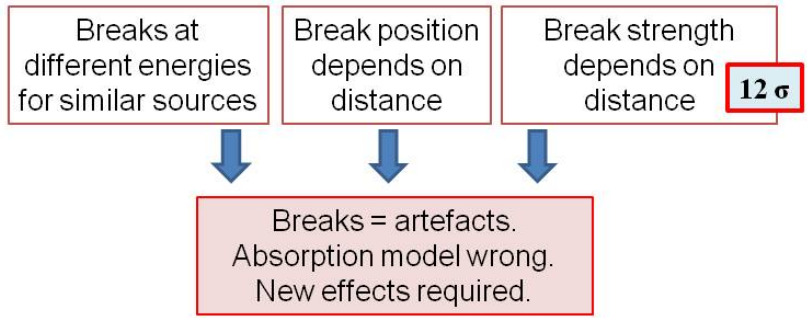

Figure 5: A schematic view of the main result of the paper.

# Supplementary information

**1. The sample construction.** The sample includes blazars observed in VHE by various instruments (see Table 1 in ED). We imposed the following requirements:

(1) the source's redshift is known and safely measured spectroscopically (we use the SIMBAD astronomical database[33] for Fermi sources and TeVCat[3] values for other sources);

(2) the mean energy of the last spectral bin $E_{\text{last}}$ after deabsorption satisfies $\log_{10}(E_{\text{last}}/E_0) > 0.1$. This last bin may have either a flux measurement or an upper limit but the previous bin is required to have a measurement;

(3) there are at least five energy bins in the measured spectrum.

For atmospheric Cerenkov telescopes, we started from the TeVCat catalog and obtained published measurements in the literature. We use only one spectrum per object and do not stack individual spectral points in order to avoid the potential variability problem.

For Fermi LAT, we preselected a sample of blazars from the 2FGL catalog[34] with measured $z \geq 0.7$, detected at $E > 10$ GeV with the test statistics $TS \geq 16$. For each of these 99 blazars we reconstruct the observed spectrum at $E > 2$ GeV in the bins (2-4) GeV, (4-10) GeV, (10-30) GeV, (30-100) GeV and (100-300) GeV with the standard *gtlike* routine from *Fermi Science Tools v9r27p1*, taking for the input the Pass 7REP (V15) data recorded from 2008 August 4 to 2014 April 19[35,36]. The lower limit of 2 GeV was chosen to cut away spectral features which often happen at slightly lower energies, so that the remaining spectrum follows a power law reasonably well. Then we applied our criterion (2) to the obtained spectra.

For the additional test (Fig. 4, ED), the selection procedure was the same except we did not put the $z \geq 0.7$ condition at preselection and replaced $E_0$ by 100 GeV in the criterion (2) at selection.

**2. Deabsorption.** For the main results of the paper, we use the lowest-absorption model available, the "fiducial" model of Gilmore et al.[5]. The optical depth for various energies and redshifts is provided by the authors of Ref.[5] at
`http://physics.ucsc.edu/~joel/EBLdata-Gilmore2012/`.

To account for the absorption, we recalculated integral number flux of photons in each energy bin. We note that the deabsorption not only increases the flux by a $\exp(\tau)$ factor but also shifts the mean energy of the bin because higher-energy photons experience stronger absorption. To account for the latter effect, we use the measured power-law spectral index for the entire spectrum we use (quoted in the referenced papers for the Cerenkov data and returned by *gtlike* for the Fermi LAT data), approximated the observed spectrum within a bin by this power law, applied the $\exp(\tau(E))$ correction to this power law and recalculated the mean energy and the integral flux in the bin for the emitted spectrum.

**3. Potential biases and systematic errors.** The first suspect to test is the Malmquist bias: faint sources would not be detected at high energies, while bright sources may have harder spectra. To address this potential bias, we included in the sample the blazars which have been detected below $E_0$ but have not been detected above (the upper limits in Fig. 3). We see that their inclusion does not affect our result, thus disfavouring this kind of a bias.

Next, there is a scatter of intrinsic properties of blazars; FSRQs are in general brighter than BLLs, hence detected farther; we see this trend also in our sample. Maybe the break position depends on the class of a blazar and just occasionally correlates with its distance, so that, for instance, all FSRQs have a break at $E \sim 100$ GeV while all BLLs have it at $E > 1$ TeV? This explanation is disfavoured by the presence, in our sample, of a very distant BLL and of a few relatively nearby FSRQs whose spectral breaks perfectly follow the general redshift dependence. A further test is given by Fig. 4 (ED) where we searched for a break at 100 GeV in a large sample of all Fermi blazars with known redshift, detected above 10 GeV. The plot suggests that the breaks appear for distant sources only, in agreement with our main result.

The errors of the energy determination for particular photons, combined with a rapidly falling spectrum, may result in an artificial overestimation of the flux in the highest-energy bin. An upward systematic error in the energy determination of all photons may shift the entire spectrum towards higher energies and, therefore, higher opacities, resulting in overestimation of the correction for the pair production. However, since we combine the data from four different instruments, one of which uses a completely different detection method, it is hardly possible that the experimental errors would sum up to a coherent effect. Though in some cases the hardening is observed in the last bin only, this is not so for a number of objects at various distances.

Another potential source of trouble is the redshift measurement: maybe they are incorrect, our sources are not that distant and we therefore overestimate the absorption? The redshift determination is a subtle task and there certainly may be erroneous values because of misidentification of spectral lines. To this end, we selected only objects with firm spectroscopic redshift measurements, so it is hardly possible that *all* their redshifts are significantly overestimated.

To summarize this discussion, it is very unlikely that the effect we see results from a selection bias or from a systematic error.